\documentclass[aps,preprint,epsfig,superscriptaddress]{revtex4-1}
\usepackage{amsmath,amssymb,graphicx}
\usepackage{latexsym}
\usepackage{esint}
\usepackage{color}
\usepackage{float}
\usepackage{multirow}
\usepackage{hyperref}

\usepackage{enumerate}

\usepackage{amsbsy}
\usepackage{bm}

\newcommand{\be}{\begin{equation}}
\newcommand{\ee}{\end{equation}}
\newcommand{\bea}{\begin{eqnarray}}
\newcommand{\eea}{\end{eqnarray}}

\newcommand{\comment}[1]{}

\begin{document}

\title{Dynamical critical behavior on the Nishimori point of frustrated Ising models}

\author{Ramgopal Agrawal$^1$, Leticia F. Cugliandolo$^{1,2}$, Lara Faoro$^3$, Lev B. Ioffe$^3$, and Marco Picco$^1$\\
{\small $^1$Sorbonne Universit\'e, Laboratoire de Physique Th\'eorique et Hautes Energies}, \\
{\small CNRS UMR 7589, 4 Place Jussieu, 75252 Paris Cedex 05, France}
\\
{\small $^2$Institut Universitaire de France, 1 rue Descartes, 75231 Paris Cedex 05, France}
\\
{\small $^3$Google Research, Mountain View, California 94043, USA}
}
\date{\today}

\begin{abstract}
By considering the \textit{quench} dynamics of two-dimensional frustrated Ising models through numerical simulations, we investigate the dynamical critical behavior on the multicritical \textit{Nishimori point} (NP). We calculate several dynamical critical exponents, namely, the relaxation exponent $z_{\rm c}$, the autocorrelation exponent $\lambda_{\rm c}$, and the persistence exponent $\theta_{\rm c}$, after a quench from the high temperature phase to the NP. We confirm their \textit{universality} with respect to the lattice geometry and bond distribution. For a quench from a power-law correlated initial state to the NP, the aging dynamics are much slower. We also look up the issue of \textit{multifractality} during the critical dynamics by investigating different moments of the spatial correlation function. We observe a \textit{single} growth law for all the length scales extracted from different moments, indicating that the equilibrium multifractality at the NP does not affect the dynamics.
\end{abstract}

\maketitle
\thispagestyle{empty}

\section{Introduction}
\label{S1}
\setcounter{page}{1} 

\textit{Critical phenomena} have been a paramount concept of statistical physics, which involves many interesting features such as \textit{universality}~\cite{stanley1971phase,goldenfeld2018lectures,cardy_1996}. The text book example of such a system is the \textit{Ising model}, where the low temperature ferromagnetic phase is separated from a high temperature paramagnetic phase via a second order phase transition point. 
The critical exponents in the pure Ising system are universal with respect to microscopic details  --- 
a property which is strongly endorsed by the renormalization group (RG) theory and many numerical works~\cite{stanley1971phase,goldenfeld2018lectures,cardy_1996,PhysRevB.30.322,PhysRevLett.75.193}. 
The universality concept is not limited to static properties as
one can also identify  \textit{dynamical} universality classes~\cite{RevModPhys.49.435,tauber_2014,RevModPhys.76.663,PhysRevE.56.2310}
which are influenced by the conservation laws imposed dynamically.
In second order phase transitions the latter are classified according to the values taken by the dynamic critical exponents. 
The idea of dynamical universality came from the application of RG methods to dynamical properties~\cite{PhysRevLett.29.1548,RevModPhys.49.435,PhysRev.177.952,10.1143/PTP.58.1142}. Two systems having the same set of static critical exponents can possess different 
dynamical exponents if, for example, they evolve with non-conserved or conserved 
order parameter dynamics. 
Therefore, in order to distinguish the universality class of a system, 
one needs both the static and dynamical critical exponents.

During the post quench \textit{critical dynamics} of an initially disordered state, 
critical fluctuations grow in the system. The extent of these fluctuations can be measured in terms of the time-dependent correlation length $\xi(t)$. A dynamical critical exponent $z_c$~\cite{RevModPhys.49.435,PhysRevB.24.1419}, which demonstrates the critical slowing down, is associated with $\xi(t)$, as $\xi(t) \sim t^{1/z_c}$. Another dynamical critical exponent $\lambda_c$ 
was found to characterize the decay of the two-time spin-spin correlations following a critical quench, i.e., $q_0(t) \sim t^{-\lambda_c/z_c}$~\cite{janssen1989new,PhysRevB.40.304,Godreche_2002,PhysRevE.72.016105} (see Sec.~\ref{S2} for the definition). Furthermore, an additional dynamical critical exponent, known as the persistence exponent $\theta_c$, was also proposed~\cite{PhysRevLett.77.3704,bray2013persistence}. This exponent was earlier studied in the context of zero-temperature coarsening dynamics~\cite{Derrida_1994,*Stauffer_1994}, and it is associated with the \textit{global} persistence probability --- the probability that the sign of the initial magnetization has not changed in time $t$ after the quench. It also decays algebraically, $p(t) \sim t^{-\theta_c}$. Therefore, there are three dynamical critical exponents, which are independent from each other (we will discuss it later). Whereas the static ones are concerned, two independent critical exponents are enough to calculate the other  ones using scaling relations. Henceforth, we need a set of five critical exponents in order to completely characterize a universality class. In pure Ising systems, all these exponents are universal (see Refs.~\cite{RevModPhys.76.663,Godreche_2002,PhysRevLett.77.3704} and references therein).

In recent years, critical disordered systems have fascinated 
many due to their application in various computational and experimental systems~\cite{nishimori2001statistical,berthier2011dynamical,PhysRevLett.103.090501}. Of particular interest are the systems with quenched randomness in terms of \textit{bond frustration}, which can dramatically alter the properties of a pure critical point. The frustration often changes the universality class of the pure critical system, e.g., new values of the critical exponents appear, and it can even induce a new thermodynamic phase, known as \textit{spin glass}. In this paper, we are primarily interested in frustrated critical systems, where the validity of universality remains controversial, e.g., see Refs.~\cite{PhysRevLett.76.4380,PhysRevLett.77.2798,Henkel_2005,PhysRevB.72.184429,PhysRevB.75.014412}. This applies to both the static and the dynamical critical exponents.

We focus on a two-dimensional $(d=2)$ frustrated Ising system, with Hamiltonian given~by
\be
\label{eq1}
H = - \sum_{\langle ij \rangle} J_{ij} S_i S_j
\; .
\ee
Here, $S_i = \pm 1$ are Ising spins, placed at each site $i$ of a $2d$ lattice. The subscript $\langle ij \rangle$ denotes the sum over all nearest neighbor (nn) spin pairs, and the exchange interaction couplings $J_{ij}$ are quenched random variables drawn from a probability distribution $P(J_{ij})$ (see below). A negative value of $J_{ij}$ indicates an anti-ferromagnetic bond (which introduces frustration), while a positive value represents a ferromagnetic bond. To examine the universality, we focus on two types of bond distributions. One is the bimodal ($\pm J$) law
\be
\label{eq2}
P(J_{ij}) = p \delta(J_{ij} + J) + (1-p) \delta(J_{ij} - J)
\; ,
\ee
where the constant $J > 0$, and $p$ is the frustration parameter ($p=0$ is the pure limit). Another probability distribution of interest is the Gaussian distribution
\be
\label{eq3}
P(J_{ij}) = \frac{1}{\sqrt{2\pi \sigma^2}}  \exp\left( -\frac{(J_{ij} - J_0)^2}{2 \sigma^2} \right)
\; ,
\ee
where $J_0$ is the mean and $\sigma^2$ is the variance of the variable $J_{ij}$. Notice that the frustration can be controlled by a parameter $r = \sigma/J_0$, with $r=0$ being the pure limit of the model~\eqref{eq1} in this case.

Due to various intriguing features like the existence of nontrivial fixed points of different universality classes including a
multicritical point and emergence of a zero-temperature spin glass phase, the model~\eqref{eq1} is quite popular in the literature~\cite{Dotsenko_1982,PhysRevB.29.4026,PhysRevLett.61.625,PhysRevLett.87.047201,Picco_2006,PhysRevB.65.054425,PhysRevB.70.134425,PhysRevE.77.051115,parisen2009strong,PhysRevE.79.021129,agrawal2023nonequilibrium}. It is also exploited to determine the error correction threshold for certain quantum error correction codes~\cite{Kitaev_1997,KITAEV20032,Iba_1999,doi:10.1063/1.1499754}. The key property which makes this model so important is the existence of a peculiar line in the phase diagram --- the \textit{Nishimori line} (see Refs.~\cite{nishimori2001statistical,Nishimori_1980,10.1143/PTP.66.1169,PhysRevLett.61.625,PhysRevB.40.9249,PhysRevB.63.104422,10.1063/1.3506843,10.1063/5.0087024} and Appendix~\ref{A1} for further details). Le Doussal and Harris~\cite{PhysRevLett.61.625,PhysRevB.40.9249} found that the multicritical point of the paramagnetic - ferromagnetic (PF) boundary is located on the intersection of the Nishimori line with the 
PF boundary; therefore, it is also known as the \textit{Nishimori point} (NP). In previous studies~\cite{PhysRevLett.87.047201,PhysRevB.79.174408}, the central charge $c$ at NP was numerically calculated, $c \simeq 0.46$. We also recently~\cite{agrawal2023nonequilibrium} found that the interfaces at this multicritical point can be uniquely described by a Stochastic Loewner Evolution (SLE$_\kappa$) with $\kappa \simeq 2.2$. The description of the critical interfaces in terms of SLE$_\kappa$ consolidates the conformal invariance of the system; also see Refs.~\cite{PhysRevLett.97.267202,PhysRevB.76.020403,PhysRevLett.100.015704} for the 
SLE$_\kappa$ description of $2d$ spin glasses and nonminimal Conformal Field Theories (CFTs).

The location of the NP in model~\eqref{eq1} has been numerically identified (see Refs.~\cite{PhysRevB.79.174408,PhysRevB.73.064410,PhysRevE.79.021129,PhysRevB.65.054425} and references therein) by the transfer matrix technique and scaling methods (see Table~\ref{tab1}). The static critical exponents at the 
NP show universality with respect to the different lattice geometries and the type of quenched disorder~\cite{PhysRevB.79.174408,PhysRevB.73.064410}. However, the universality of the dynamical critical exponents at the NP has not been investigated yet. Moreover, the values of the different dynamical critical exponents are not known except for $z_c$~\cite{Ozeki_1998,agrawal2023nonequilibrium}. It will be interesting to see whether the dynamical critical exponents maintain the universality, or if not, establish the way in which they may vary. To answer these questions, we consider various lattices of different geometries, namely, the square lattice (SL), triangular lattice (TL), and honeycomb lattice (HL). On each of these lattices, we introduce bimodal and Gaussian bond disorders. Therefore, we have six different systems with a unique lattice geometry and quench disorder. In order to calculate the dynamical critical exponents, we quench the system initially kept at infinitely high temperature to the NP. We also investigate the quench dynamics after a start from a power-law correlated state, i.e., a critical Ising configuration. For all these purposes, we exploit single-spin-flip Monte Carlo methods with non-conserved order parameter.

Another interesting feature we examine is the \textit{multifractal} nature of time-dependent correlation functions. At a pure critical point, different moments of the equilibrium spatial correlation function $C_q(r)$ decay as $C_q(r) \sim 1/r^{\eta_q}$, with $\eta_q = q \eta_1$. The latter relation ($\eta_q$ being an integer multiple of $\eta_1$) indicates the usual monofractal behavior of different moments. However, at the disorder induced transition points in random systems the moments generally exhibit a multifractal behavior~\cite{LUDWIG1990639,Lewis_1998,PhysRevB.60.3428,palagyi2000boundary,DAVIS2000713,Monthus_2009,marinari2023multiscaling}, as the magnetic exponents $\eta_q \ne q \eta_1$. This multifractality also holds at the NP~\cite{PhysRevLett.87.047201,PhysRevB.65.054425} --- on the Nishimori line the moments are equal two by two, with the exponents $\eta_q \ne q \eta_1$ additionally at NP (see Sec.~\ref{S3} ahead for more details). In this paper, we examine whether this \textit{static} multifractality affects the critical dynamics of the NP. We extract the growing length scales from different time-dependent moments $C_q(r,t)$ and we investigate whether they possess a single growth law akin to the average correlation length $\xi(t)$, or if the dynamics is dominated by multiple growth laws.

The main findings of our paper are as follows:
\vspace{-0.35cm}
\begin{itemize}
\item[(1)]~The dynamical critical exponents $z_{\rm c}$, $\lambda_{\rm c}$, and $\theta_{\rm c}$ 
 are universal with respect to the lattice geometry and bond distribution. 
 \vspace{-0.35cm}
\item[(2)]~For quenches from the Ising critical point to the NP aging is much slower than for quenches from fully disordered
 initial states, with $q_0(t) \sim t^{- 0.03}$. 
 \vspace{-0.35cm}
\item[(3)]~A single growth law is observed for all the length scales extracted from different moments $C_q(r,t)$, i.e., static multifractality does not affect the post-quench critical dynamics.
\end{itemize}
\vspace{-0.22cm}

We have structured this paper in the following manner. In Sec.~\ref{S2}, we detail the numerical techniques and the definitions of various observables later required in the paper. In Sec.~\ref{S3}, we present our main numerical results. Finally, in Sec.~\ref{S4}, we summarize and discuss our results. 
Appendix~\ref{A1} discusses some basic details of the Nishimori line.

\section{Numerical techniques and definitions}
\label{S2}

We investigate the nonequilibrium dynamics of the $2d$ Ising model on three types of lattice geometries (SL, TL, and HL) considering periodic boundary conditions (PBCs) in both lattice directions. The coordination numbers of these lattices are $n_{\rm SL}=4$, $n_{\rm TL}=6$ and $n_{\rm HL}=3$, respectively. In order to circumvent the complexity in the construction of nontrivial lattices, we use the following trick discussed in Ref.~\cite{Blanchard_2017}. The TL is built by adding a diagonal bond between the $(i,j)th$ and $(i+1,j+1)th$ lattice sites. For the HL, we remove the bonds between all $(i,j)th$ and $(i,j+1)th$ sites if $i+j$ is an even number. If $i+j$ is an odd number, the bonds between the $(i,j)th$ and $(i,j-1)th$ sites are removed. In this way, we always have $N=L^2$ number of spins in the lattice. The pure $2d$ Ising model has a second order critical point at temperatures $T_{\rm Is}^{\rm SL} \simeq 2.269$, $T_{\rm Is}^{\rm TL} \simeq 3.641$, and $T_{\rm Is}^{\rm HL} \simeq 1.518$ (in units of $J/k_{\rm B}$; $k_{\rm B}$ being Boltzmann constant).

\begin{table}[t!]
	\begin{center}
		\begin{tabular}{ c  c  c }
			\hline
			\hline
			\text{Lattice} & $\pm J$ & $G$ \\
			\hline
			\text{SL} & $T_{\rm N} = 0.953(2)$ & $T_{\rm N} = 0.958(1)$ \\

			\text{TL} & $T_{\rm N} = 1.228(1)$ & $T_{\rm N} = 1.254(2)$ \\

			\text{HL} & $T_{\rm N} = 0.759(1)$ & $T_{\rm N} = 0.788(1)$ \\
			\hline
			\hline
		\end{tabular}
	\end{center}
	\caption{Values of the critical temperature $T_{\rm N}$ at the Nishimori point (NP) in the $2d$ frustrated Ising model on the square lattice (SL), triangular lattice (TL), and honeycomb lattice (HL), with bimodal ($\pm J$) and Gaussian $(G)$ bond distributions. These values are quoted from earlier studies~\cite{PhysRevB.79.174408,PhysRevB.73.064410,PhysRevE.79.021129,PhysRevB.65.054425} and references therein. The corresponding error bars account for the range of $T_{\rm N}$ considered for the localization estimates. The frustration parameter ($p_{\rm N}$, or $r_{\rm N}$) can be calculated using the relations in Eq.~\eqref{eq6}.}
	\label{tab1}
\end{table}

At time $t=0$, the spin configuration is prepared in a high temperature paramagnetic phase by assigning 
random values $(\pm 1)$ with probability a half to each spin $S_i$. The system is then quenched to the NP. 
The values of the Nishimori critical temperatures for different lattices and bond disorders are mentioned in 
Table~\ref{tab1}. Since the model~\eqref{eq1} does not have any intrinsic dynamics, we exploit the \textit{Glauber-nonconserved-kinetics} method where the lattice system is connected with a giant heat bath. This heath bath introduces 
single spin flip dynamics. Furthermore, in order to ensure detailed balance, 
we flip a single spin with the Metropolis transition rate~\cite{metropolis1949monte,newman1999monte}
\be
W(S_i\to -S_i) = N^{-1}\min \left\{1,e^{-\frac{\Delta E}{T_c}}\right \}
\; ,
\label{metrop}
\ee
where $\Delta E$ is the energy difference in the proposed spin flip. Time is measured in terms of Monte Carlo 
Steps (MCS), each corresponding to $N = L^2$ attempted spin flips. The Boltzmann constant is set to unity without the loss of any generality.

When a spin system is thermally quenched to a critical point, the initial state 
becomes unstable and the system prefers to reach the new critical equilibrium. 
In order to minimize the net free energy, the spins locally group as in the target critical state --- resulting in the growth of (spatial) correlations with time. To investigate these correlations quantitatively, the standard 
practice is to consider the two-point spatial correlation function
\be
\label{corr}
C(r,t) = \frac{1}{N} \sum_{i} \overline{ \langle  S_i(t) S_{i+\vec r}(t) \rangle }
\; ,
\ee
where $\langle (\cdots) \rangle$ denotes an average over different random initial conditions and thermal noise, while $\overline{ (\cdots) }$ denotes the average over disorder realizations. For a quench from the paramagnetic phase, the above time-dependent correlation function respects the following scaling relation
\be
\label{scale}
C(r,t) = \frac{1}{r^{\eta}} \; \overline F\left(\frac{r}{\xi (t)}\right) 
\; ,
\ee
where $\overline{F}(s)$ is a scaling function with $\overline{F}(0) = 1$, $\eta$ is the standard critical exponent, and $\xi (t)$ is the dynamical correlation length. For the calculation of $\xi(t)$, one can assume an exponential form of the function 
$F(r,t) ~[=r^{\eta} C(r,t)] \sim {\rm e}^{-r/\xi(t)}$, where $\xi(t)$ tends to $\xi_{\rm eq}$ as $t \rightarrow \infty$. Another widespread method to extract $\xi (t)$ is to fix the fall of the function $F(r,t)$ as, $F(r=\xi (t),t) = F_0$. In this paper, we choose $F_0 \equiv 1/{\rm e}$.

During the post quench dynamics, the time-dependent correlation length $\xi(t)$ grows in a power-law fashion,
\be
\label{length}
\xi(t) \sim t^{1/z_c}
\; ,
\ee
where $z_c$ is a dynamical critical exponent, known as the relaxation exponent. It describes the critical slowing down near the transition point. In the pure $2d$ Ising model, the universal value of $z_c$ is around $z_c \simeq 2.17$, see Refs.~\cite{PhysRevE.56.2310,PhysRevB.62.1089,Ricateau_2018}.

The time correlations during the approach to a critical point are also crucial. Especially, the two-time correlations have been shown to exhibit universal behavior in terms of \textit{critical aging}. We mention that the aging phenomenon is a 
widely studied concept in the non-equilibrium dynamics of disordered systems, e.g., spin-glasses and 
random-field systems; see Refs.~\cite{CuKu93,doi:10.1142/9789812819437_0006,henkel2011non,PhysRevB.49.6331,PhysRevB.91.174433,PhysRevE.104.044123,PhysRevE.108.044131}. To calculate the associated critical exponent, one can simply study the two-time correlations from the overlap of a configuration at time $t$ with its initial configuration,
\be
\label{auto22}
q_0(t) = \frac{1}{N} \sum_{i} \overline{ \langle  S_i(0) S_i(t) \rangle }
\; .
\ee
During the quench dynamics from a high temperature phase to the critical point, the above quantity has algebraic behavior,
\be
\label{auto2}
q_0(t) \sim t^{- \lambda_c/z_c}
\; .
\ee
Here, $\lambda_c$ is known as the autocorrelation exponent, which is another independent dynamical critical exponent~\cite{janssen1989new}. Huse~\cite{PhysRevB.40.304} has obtained a precise estimation of $\lambda_c$ ($\lambda_c \simeq 1.59$) using Monte Carlo simulations for the pure $2d$ Ising model with SL geometry. As a benchmark, we have separately confirmed that this value of $\lambda_c$ is universal with respect to the different lattice geometries concerned in this paper.

To investigate the aging process, the above two-time function is generalized as
\be
\label{auto1}
C(t,t_w) = \frac{1}{N} \sum_{i} \overline{ \langle  S_i(t) S_i(t_w) \rangle }
\; ,
\ee
where $t_w (\ne 0) < t$ is the waiting time, which is also known as the \textit{age} of the system. When $t - t_w \ll t_w$, the above expression respects the time-translation invariance (TTI), and  the fluctuation dissipation theorem (FDT) also holds. However, for large time separations $t - t_w \gg t_w$, TTI is no longer respected and FDT is also broken down. In the latter regime, the autocorrelation function~\eqref{auto1} has the following scaling form~\cite{Godreche_2002,henkel2011non}
\be
\label{auto3}
C(t,t_w) \sim t_{w}^{-\eta/ z_c} \left( \frac{t}{t_w} \right)^{- \lambda_c/z_c}
\; .
\ee
The above relation can be written in terms of the ratio of time-dependent correlation lengths~as
\be
\label{auto4}
C(t,t_w) \sim \xi(t_{w})^{-\eta} \left( \frac{\xi(t)}{\xi(t_w)} \right)^{- \lambda_c}
\; ,
\ee
which in the asymptotic regime $\xi(t) \gg \xi(t_w)$ behaves as $C(t,t_w) \sim [\xi(t)]^{- \lambda_c}$.

Furthermore, the global persistence probability $p(t)$ is defined as,
\be
\label{pers1}
p(t) = 1 - \sum_{{t'}=0}^{t'=t} {\cal N}(t')
\; ,
\ee
where ${\cal N}(t)$ denotes the fraction of systems which have flipped their sign of magnetization at time-instant $t'$, for the
first time, after the quench at the initial time $t' = 0$. Therefore, $p(t)$ simply gives the probability that the sign of the 
magnetization has not changed in time $t$ after the quench. During the critical quench dynamics from a high temperature state, $p(t)$ falls in power-law manner~\cite{PhysRevLett.77.3704,bray2013persistence} as,
\be
\label{pers2}
p(t) = t^{-\theta_c}
\; ,
\ee
where $\theta_c$ is the persistence exponent. As the net magnetization in a paramagnetic state and also at a critical point is zero, during simulations, some nonzero magnetization density [say, $M(0)$] is assigned by hand to the initial state. Notice that $p(t)$ is the probability that the sign of the global magnetization is unchanged after a time $t$ during the dynamics. The above behavior~\eqref{pers2} is obtained in the limit of $M(0) \rightarrow 0$.

During the dynamics, as long $L \gg \xi(t)$, the time-dependent magnetization density $M(t)$ can be treated as a Gaussian process. If one additionally assumes that $M(t)$ is a Gaussian \textit{Markov} process, a scaling relation~\cite{PhysRevLett.77.3704} for the persistence exponent $\theta_c$ is given by
\be
\label{pers3}
\theta_c z_c = \lambda_c -d + 1 -\eta/2
\; .
\ee
We remark that in the pure Ising model~\cite{PhysRevLett.77.3704} and in the diluted Ising model~\cite{Paul_2005}, the above relation breaks down, respectively, in the two-loop and one-loop order of the $d_{\epsilon} = 4 - \epsilon$ dimensional expansion. This led people to believe that $M(t)$ is rather a Non-Markovian process, and the exponent $\theta_c$ is an independent dynamical critical exponent. In the following, we will examine whether the same holds for the present frustrated system.

\begin{figure}[t!]
	\centering
	\rotatebox{0}{\resizebox{.85\textwidth}{!}{\includegraphics{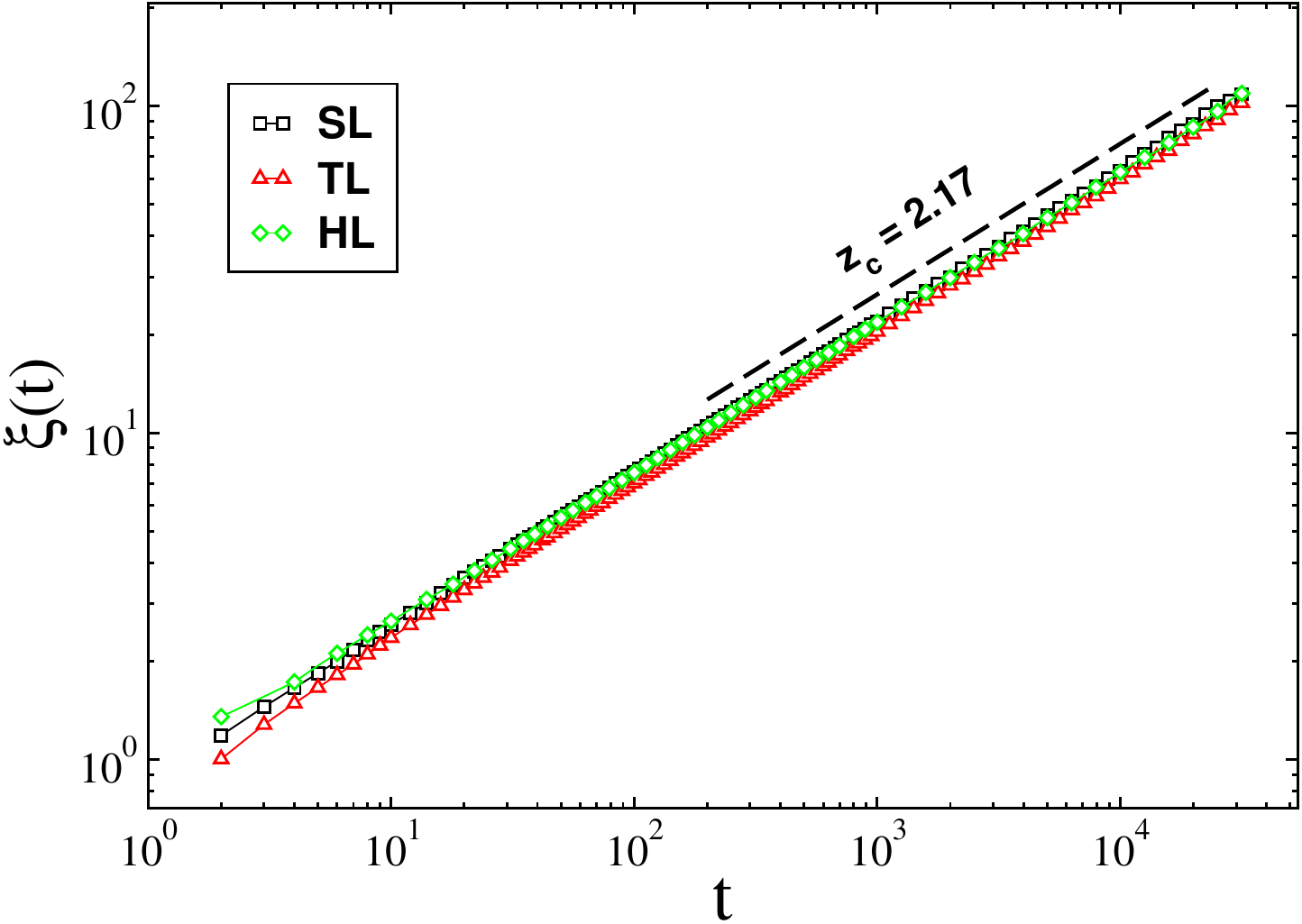}}}
	\caption{Plot of the correlation length $\xi (t)$ vs time $t$, in log-log scale, 
	for critical Ising quenches, that is to  $T_{\rm Is}$, of systems with different lattice geometries (see the key) 
	from infinitely high temperature $T \gg T_{\rm Is}$ initial conditions. 
	The linear size of each system is fixed to $L = 512$. The dashed line indicates the pure growth law $\xi (t) \sim t^{1/z_{\rm c}}$, with $z_{\rm c} = 2.17$.}
	\label{fig1}
\end{figure}

\begin{figure}[t!]
	\centering
	\rotatebox{0}{\resizebox{.85\textwidth}{!}{\includegraphics{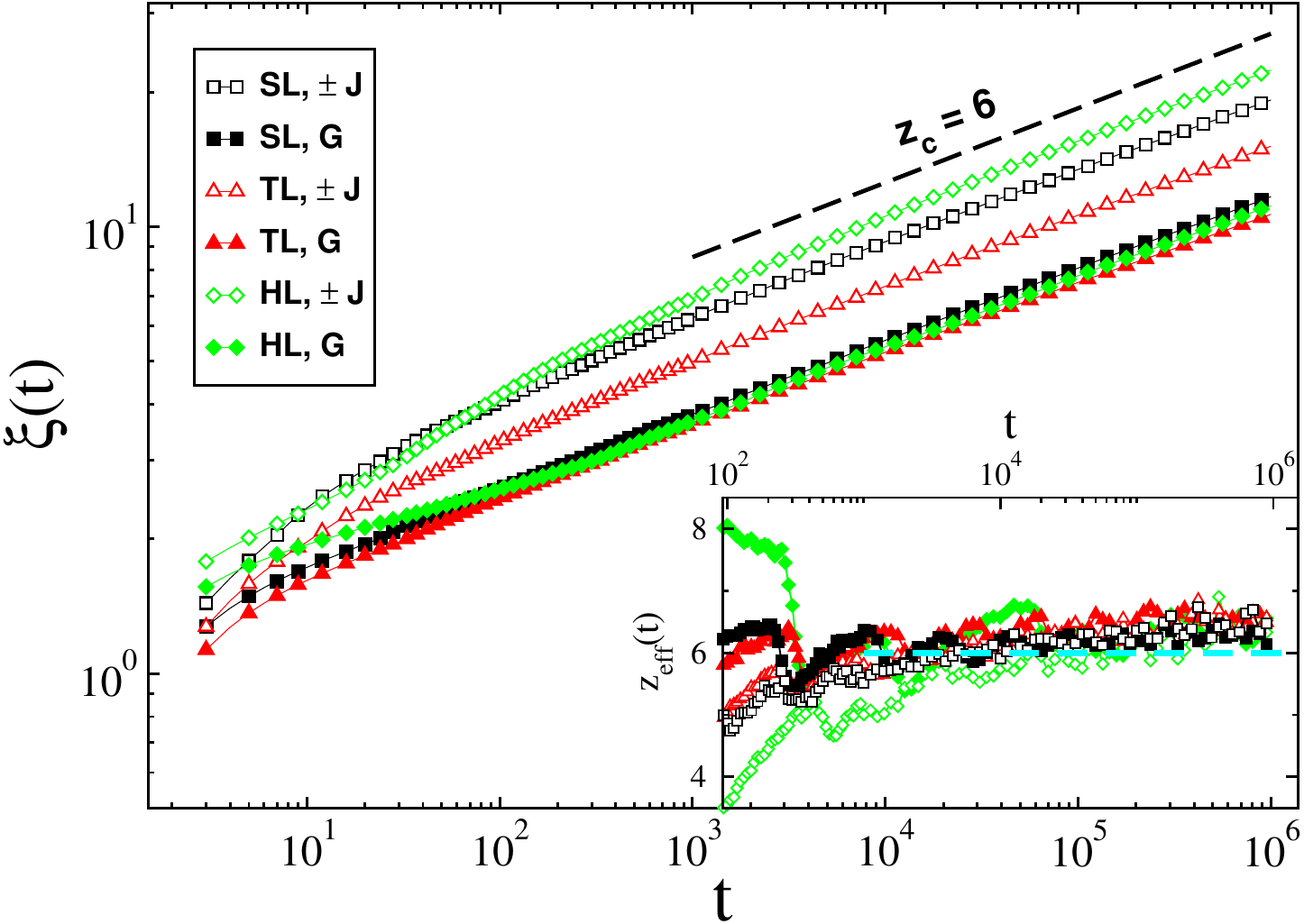}}}
	\caption{Plot of the correlation length $\xi (t)$ vs time $t$, in log-log scale, for a quench to the 
	Nishimori point $T_{\rm N}$ from infinitely high temperature $T \gg T_{\rm N}$ initial conditions. 
	Different datasets belong to systems with different lattice geometries and distributions of bond disorder (see the key). For each system, the linear size is $L = 512$. The dashed line represents the law $\xi (t) \sim t^{1/z_{\rm c}}$, with $z_{\rm c} = 6$. In the inset, 
	the effective exponent $z_{\rm eff}(t)$ is plotted against $t$ for the datasets in the main frame. 
	The dashed horizontal line indicates the asymptotic dynamical exponent $z_{\rm c} = 6$.}
	\label{fig2}
\end{figure}

\section{Main results}
\label{S3}

In the present section, we discuss our main results for the $2d$ frustrated Ising model defined on different lattice geometries with different quenched bond disorders (mentioned above).

\begin{table}[t!]
	\begin{center}
		\begin{tabular}{ c  c  c  c}
			\hline
			\hline
			\text{system} & $z_{\rm c}$ & $\lambda_{\rm c}/z_{\rm c}$ & $\theta_{\rm c}$ \\
			\hline
			SL, $\pm J$ & $6.31(5)$ & $0.224(3)$ & $0.083(2)$ \\
			
			SL, $G$ & $6.22(3)$ & $0.21(1)$ & $0.077(5)$\\
			
			TL, $\pm J$ & $6.39(4)$ & $0.219(4)$ & $0.082(3)$\\
			
			TL, $G$ & $6.49(3)$ & $0.217(3)$ & $0.081(2)$\\
			
			HL, $\pm J$ & $6.18(5)$ & $0.222(5)$ & $0.084(2)$\\
			
			HL, $G$ & $6.35(4)$ & $0.224(2)$ & $0.085(1)$\\
			\hline
			\hline
		\end{tabular}
	\end{center}
	\caption{Values of the exponents $z_{\rm c}$, $\lambda_{\rm c}/z_{\rm c}$ and $\theta_{\rm c}$, for a quench to the Nishimori point (NP) at $T = T_{\rm N}$ from a high temperature ($T \gg T_{\rm N}$) state, in the $2d$ frustrated Ising model defined on different geometries [square lattice (SL), triangular lattice (TL), and honeycomb lattice (HL)] with different bond distributions [bimodal ($\pm J$) and Gaussian ($G$)]. The values and corresponding (statistical) error bars are estimated from the best power-law fits to the long time behavior of the quantities $\xi (t)$, $q_0(t)$ and $p(t)$, with the \textit{reduced} $\chi^2$-value being near $1$.}
	\label{tab2}
\end{table}

\subsection{Growth of the dynamical correlation length}

We first benchmark the known growth law $\xi(t) \sim t^{1/z_{\rm c}}$ (with $z_{\rm c} \simeq 2.17$) for a pure system. 
For this, we perform a quench at time $t=0$ from an initial paramagnetic state to the Ising critical point $T = T_{\rm Is}$. 
In Fig.~\ref{fig1}, the correlation length $\xi(t)$ is plotted against time $t$ for various systems having 
different lattice geometries. The linear size of each system is fixed to $L = 512$. The data presented in this figure are 
averaged over $1000$ independent runs. As expected, different datasets confirm the power law growth with exponent $z_{\rm c} \simeq 2.17$.

Now, let us discuss the results for a quench to the NP ($T = T_{\rm N}$) from an infinitely high temperature $T \gg T_{\rm N}$. In the main frame of Fig.~\ref{fig2}, the correlation length $\xi(t)$ is plotted against $t$ for systems with different geometries and bond distributions. The linear size of each system is again $L = 512$, with a non-equilibrium ensemble average performed over $1000$ independent initial configurations and disorder realizations.

In the late time regime, all the datasets exhibit power law behavior of the form $\xi(t) \sim t^{1/z_{\rm c}}$ over a three decades of time, on the log-log scale of the plot. Such a promising behavior spread over multiple decades should correspond to the asymptotic limit. To estimate the value of the exponent $z_{\rm c}$ from a dataset, we performed various power-law fits by varying the time-window within the temporal regime $t\in[10^4,10^6]$. The fits with the values of \textit{reduced}-$\chi^2$ near $1$ are accepted, and the corresponding estimates of the exponent $z_{\rm c}$ are mentioned in Table~\ref{tab2}. The statistical error bars are obtained using \textit{Jackknife} resampling technique~\cite{doi:10.1137/1.9781611970319}, which also accounts for the correlations in a dataset. However, we do not take these error bars seriously as they do not account for the systematic errors explained below.

Apart from the statistical errors reported in Table~\ref{tab2} the fits may also have systematic errors, which are often harder to capture. For this purpose, we first employ different ways of calculating $\xi(t)$. For all the systems reported in Table~\ref{tab2} the exponent $z_{\rm c}$ varies significantly beyond the statistical error bars; however, we always obtained $z_{\rm c}$ between $6.0 - 6.5$. Furthermore, the localization of the NP also ensues the errors in $z_{\rm c}$, as the values of $T_{\rm N}$ reported in Table~\ref{tab1} are just numerical estimates. To account for these errors, we vary the value of $T_{\rm N}$ by about $0.01$. We find again that the exponent $z_{\rm c}$ changes but remains within the range $6.0 - 6.5$.

To gain some more insight, let us also look at the effective exponent $z_{\rm eff}(t)$ defined as
\be
\label{effExp}
\frac{1}{z_{\rm eff}(t)} = \frac{d\ln \xi(t)}{d\ln t}
\; ,
\ee
with $\lim_{t\rightarrow \infty} z_{\rm eff}(t) = z_{\rm c}$. We determine the above derivative at discrete simulation times using the central-difference-scheme as,
\be
\label{effExp1}
\frac{1}{z_{\rm eff}(t_i)} =\frac{ \ln \xi(t_{i+1})- \ln \xi(t_{i-1}) }{ \ln(t_{i+1}) - \ln(t_{i-1}) }
\; .
\ee
In the inset of Fig.~\ref{fig2}, $z_{\rm eff}(t)$ is plotted against $t$ for different datasets in the main frame. 
The asymptotic value of $z_{\rm eff}(t)$ for different lattices and bond distributions fluctuates between $6.0-6.5$. However, the data in the inset also show that the fluctuations are of oscillatory nature, 
which eventually should die out in a longer simulation. Furthermore, for an independent check of the correlation length $\xi(t)$, we analyze the fluctuations $\Delta M(t) =\overline{\langle M^2 \rangle} - \overline{\langle M \rangle}^2$ in the magnetization density $M(t)$. According to the short-time critical dynamics (STCD) approach~\cite{PhysRevE.58.4242,*Albano_2011}, this quantity behaves as $\Delta M(t) \sim t^{\zeta}$, with $\zeta = (d -\eta)/z_c$ after a quench from infinitely high temperature. This analysis gives a value of $z_{\rm c}$ between $6.1 - 6.6$ for the different systems considered in this work (data not shown here).

In conclusion, the value of $z_{\rm c}$ is  \textit{universal} with respect to different lattice geometries and bond distributions, but finding a precise value solely from the numerical simulations is hard. We expect that in the long time limit (and for large system sizes) a universal value of $z_c \simeq 6$~\cite{agrawal2023nonequilibrium} will be approached. 
See Sec.~\ref{S4} for a more detailed discussion.

\begin{figure}[t!]
	\centering
	\rotatebox{0}{\resizebox{.85\textwidth}{!}{\includegraphics{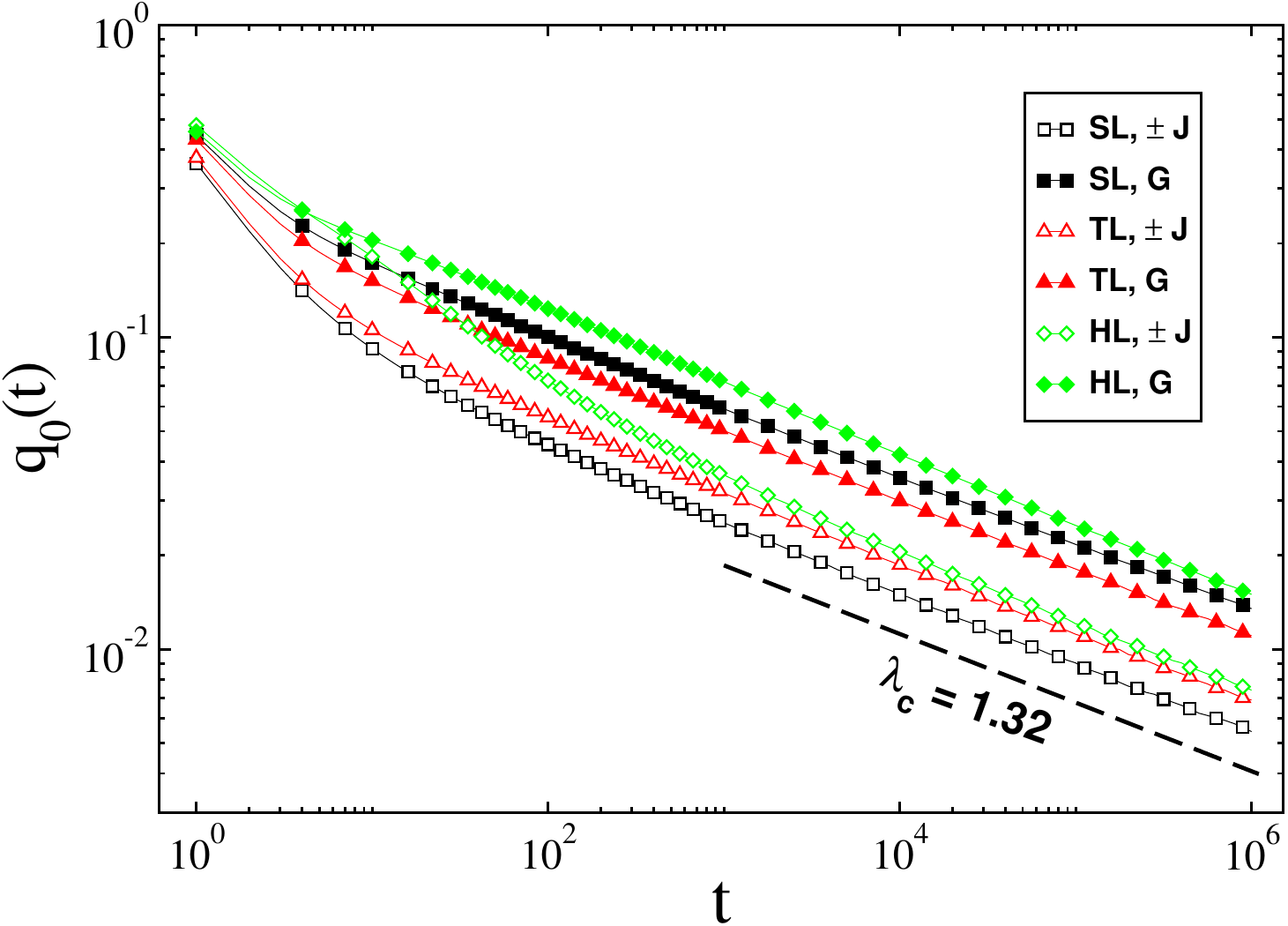}}}
	\caption{Plot of the correlation $q_0(t)$ vs time $t$, in log-log scale, 
	for a quench to the Nishimori point $T_{\rm N}$ from infinitely high temperature $T \gg T_{\rm N}$ initial conditions. 
	Different datasets belong to systems with different lattice geometries and distributions of bond disorder (see the key). The dashed line denotes the power law $q_0(t) \sim t^{-\lambda_c/z_{\rm c}}$, with $\lambda_c = 1.32$ and $z_{\rm c} = 6$.}
	\label{fig3}
\end{figure}

\begin{figure}[t!]
	\centering
	\rotatebox{0}{\resizebox{.85\textwidth}{!}{\includegraphics{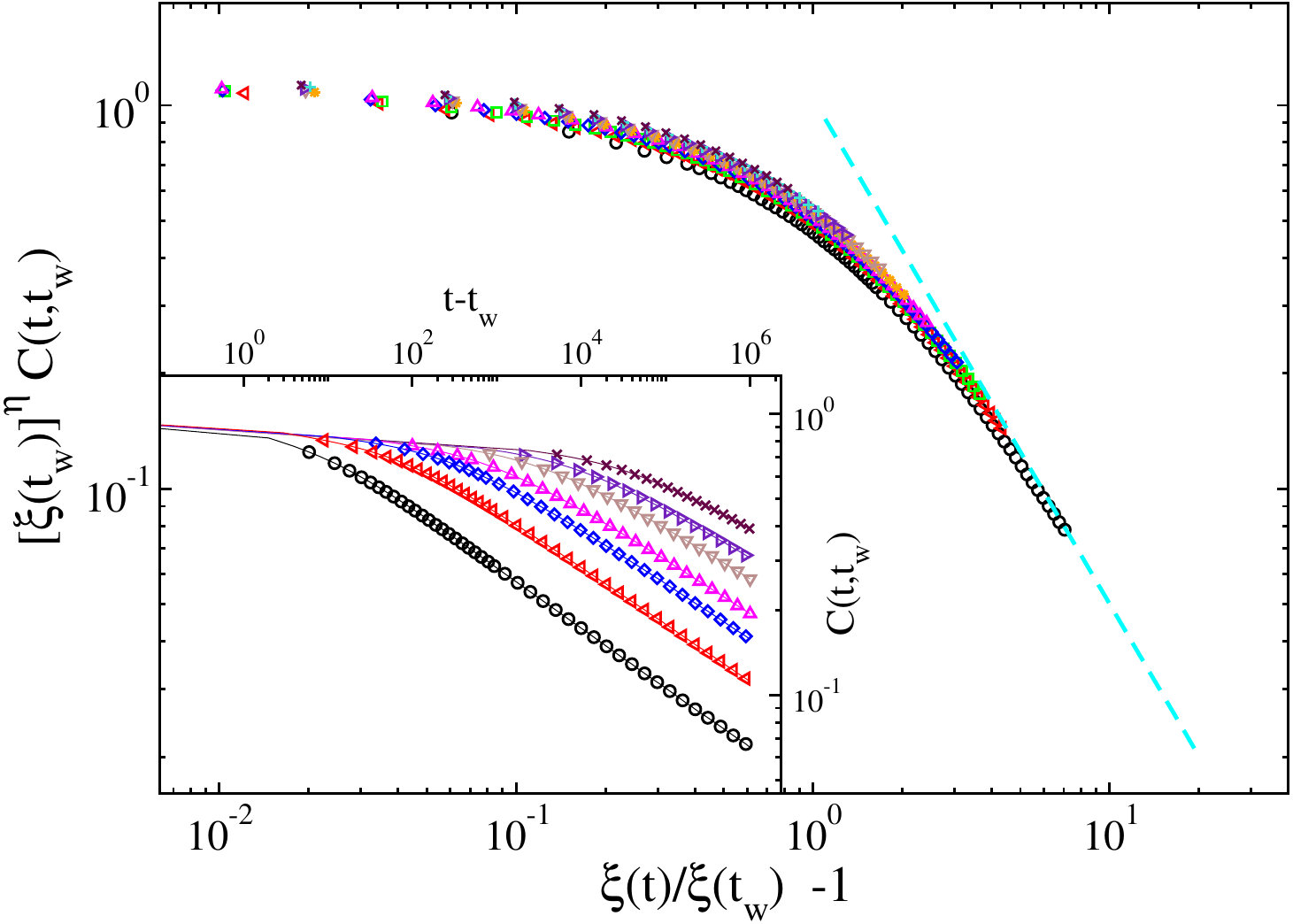}}}
	\caption{Log-log plot of the scaling variable $\xi(t_{w})^{\eta} C(t,t_w)$ against $\xi(t)/ \xi(t_w) - 1$ of 
	systems defined on the square lattice with bimodal $(\pm J)$ disorder. 
	Quench to the Nishimori point $T_{\rm N}$ from infinite high temperature $T \gg T_{\rm N}$ initial conditions. 
	Different datasets correspond to waiting times $t_w \in [10,20 000]$. The dashed line represents the asymptotic behavior $f(y) \sim y^{-\lambda_c}$, with $\lambda_c = 1.32$. Inset: Plot of $C(t,t_w)$ vs time difference $t-t_w$ for the same 
	data used in the main frame, also in log-log scale. $t_w$ increases from bottom to top.}
	\label{fig4}
\end{figure}

\subsection{Critical aging}

Now let us understand the aging phenomenon during the critical dynamics on the  NP. We first present the results for a quench from the paramagnetic state ($T \gg T_{\rm N}$) to $T = T_{\rm N}$. This set up follows the zero-field-cooling (ZFC) protocol, but with the initial temperature $T = \infty$.

In Fig.~\ref{fig3}, the two-time spin correlation $q_0(t)$ is plotted against $t$ for different systems mentioned above (the size of the system and the ensemble averages are similar to the ones in the previous subsection). Asymptotically, the quantity $q_0(t)$ falls in a power-law fashion~\eqref{auto2}, which is clearly observed for all the datasets in Fig.~\ref{fig3}. Again, to become sensitive, we performed power-law fits in the late time regimes of different datasets. Observed best-fit values of the exponent $\lambda_{\rm c}/z_{\rm c}$ are listed in Table~\ref{tab2}. The value of $\lambda_{\rm c}/z_{\rm c}$ is around $0.22$ for all datasets (within error bars), which gives a universal value of $\lambda_{\rm c} \simeq 1.32$, when using $z_{\rm c} \simeq 6$.

We now discuss the generalized two-time correlation function $C(t,t_w)$ \eqref{auto1} at nonzero waiting times $t_w$. As discussed in the previous section, this quantity exhibits aging for $t - t_w \gg t_w$. For brevity, we show in  Fig.~\ref{fig4} the results only for a SL system with $\pm J$ disorder. Let us first understand the inset of this figure, where $C(t,t_w)$ is plotted against time-difference $t - t_w$ for different $t_w \in [10, 20 000]$. When $t - t_w \sim t_w$, $C(t,t_w)$ does not change a lot and exhibits a plateau. However, when $t - t_w \gg t_w$, $C(t,t_w)$ falls in a power-law fashion, with no collapse for datasets with different $t_w$. The latter indicates a clear violation of TTI. The aging can be seen from the fact that $C(t,t_w)$ falls to a fixed value, say $C_0$, more quickly for datasets with smaller $t_w$.

In the main frame of Fig.~\ref{fig4}, we test the scaling function~\eqref{auto4}, for which we plot the scaling variable $\xi(t_{w})^{\eta} C(t,t_w)$ against $\xi(t)/ \xi(t_w) - 1$. Datasets with different $t_w$ appear to collapse on each other, confirming the relation~\eqref{auto4}. Notice that the data in this figure are not in the \textit{true} asymptotic regime (see $\text{max}[\xi(t)/ \xi(t_w)]$). Therefore, the behavior $C(t,t_w) \sim [\xi(t)]^{-\lambda_{\rm c}}$; with $\lambda_{\rm c} \simeq 1.32$ (indicated by a dashed line in figure), is not observed in the current time scales. However, the approach towards this behavior is clearly seen at large values of $\xi(t)$.

Let us also check what happens to the aging properties if the system is quenched from an Ising critical point 
($T = T_{\rm Is}$) to the NP ($T = T_{\rm N}$). We remark that the spin configuration in 
a critical Ising state is \textit{power-law} correlated,
\be
\langle S_i(0) S_{i+\vec{r}}(0) \rangle \propto \frac{1}{r^{\eta_{\rm Is}}}
\label{ising}
\; .
\ee
Here, the critical exponent is $\eta_{\rm Is} = 1/4$. The target spin configuration at the NP 
is also power-law correlated but with a different exponent $\eta \simeq 0.18$~\cite{PhysRevLett.87.047201}, 
i.e., the decay in the target state is \textit{comparatively} slower.

\begin{figure}[t!]
	\centering
	\rotatebox{0}{\resizebox{.85\textwidth}{!}{\includegraphics{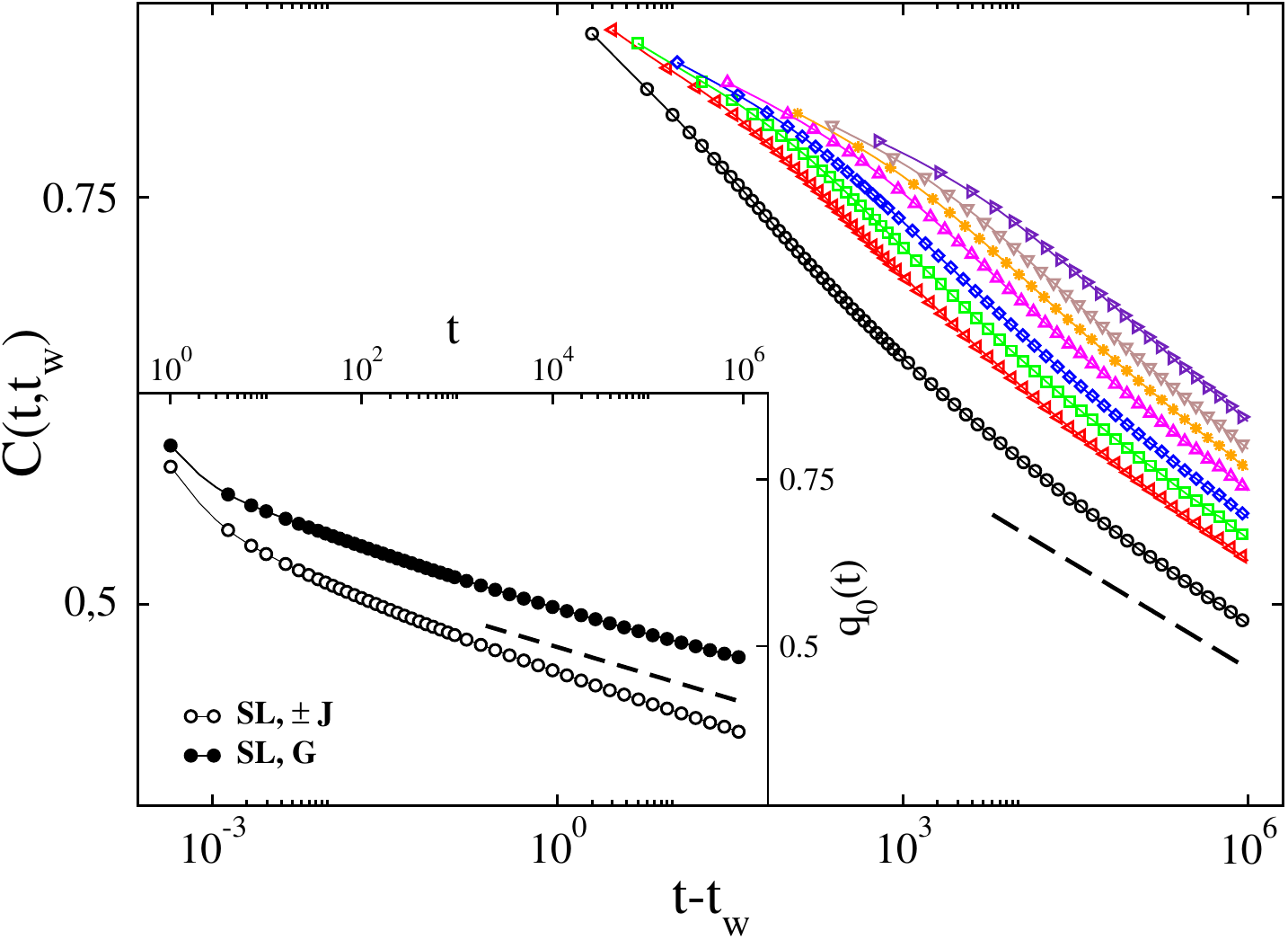}}}
	\caption{Main frame: Log-log plot of the autocorrelation function $C(t,t_w)$ against time difference 
	$t-t_w$ of a system defined on the square lattice (SL) with bimodal $(\pm J)$ disorder, for a quench 
	from the Ising critical point $T_{\rm Is}$ to the Nishimori point $T_{\rm N}$. Different datasets correspond 
	to waiting times $t_w \in [10,10 000]$, where $t_w$ increases from bottom to top. Inset: Plot of $q_0(t)$ vs $t$, also in log-log scale, for a quench from $T_{\rm Is}$ to $T_{\rm N}$ of SL systems with $\pm J$ and Gaussian ($G$) bond distributions (see the key). The dashed lines in both plots denote the expected asymptotic behavior $f(t) \sim t^{-\eta/z_c}$, with $\eta = 0.18$ and $z_c = 6$.}
	\label{fig5}
\end{figure}

We mention for the convenience of the reader that the aging during the quenches from an initial correlated state has earlier been studied by Bray and co-workers~\cite{PhysRevB.43.3699,Humayun_1991}. They found for pure Ising 
systems that there is a threshold value of the decay exponent ($\eta_{\rm th} \simeq 3/2$ in $d=2$) below which the initial correlations are relevant in the RG sense. In our present case the situation is more complicated as the target state is also correlated. We expect that the effect of initial correlations shall persist during the critical dynamics. However, as $\eta_{\rm Is}$ is larger than the value of $\eta$ at the NP, some crossover is expected. With these questions in mind, we present data for 
the two-time correlations in Fig.~\ref{fig5}. In the inset, $q_0(t)$ is plotted against $t$ for two different systems (SL with $\pm J$ and Gaussian disorders). On the log-log scale of the plot, a clean power-law behavior is observed for both datasets at timescales $t > 10^3$. As expected~\footnote{For a quench from $T_{\rm Is}$ to $T < T_{\rm Is}$, a decay exponent around $\eta_{\rm Is}/4$ was observed~\cite{PhysRevB.43.3699,Humayun_1991}.}, the decay 
is much slower (exponent around $0.03$) than that for a quench from high temperature state (where the exponent 
is $\lambda_{\rm c}/z_{\rm c} \simeq 0.22$). This is surely a \textit{persisting} effect of the initial correlated state. More quantitative reasoning behind this new exponent is obtained below.

We now look at the two-time correlation for nonzero waiting times $t_w$. In the main frame of Fig.~\ref{fig5}, $C(t,t_w)$ is plotted against time-difference $t-t_w$ for various $t_w$. For brevity, the presented data are only for one system (SL with $\pm J$-disorder). For large time differences, no scaling collapse is observed in such kind of plot. Moreover, a clean aging signature is observed from the fall of different  $t_w$-datasets. At late time scale (the onset of which increases with $t_w$), all datasets indicate a power-law fall with the above observed exponent $\sim 0.03$. Therefore, this value around $0.03$ should have some deeper meaning explained as follows. Due to the initial relevant correlations, the loss of memory during the dynamics is suppressed -- resulting in a smaller value of the aging exponent. However, as the target state is also power-law correlated ($\eta \simeq 0.18 < \eta_{\rm Is}$), a behavior akin to that of a quench from a high temperature state (exponent $\lambda_{\rm c}/z_{\rm c} \simeq 0.22$) should be obtained at larger time scales (beyond the ones currently presented). Notice that even for quenches from high $T$, $C(t,t_w) \sim t^{-\eta/z_{\rm c}}$ for $t \simeq t_w$~\eqref{auto3}. In the present scenario, the nontrivial exponent $0.03$ is in agreement with $\eta/z_{\rm c}$ (where $\eta \simeq 0.18$ and $z_{\rm c} \simeq 6$).

\begin{figure}[t!]
	\centering
	\rotatebox{0}{\resizebox{.85\textwidth}{!}{\includegraphics{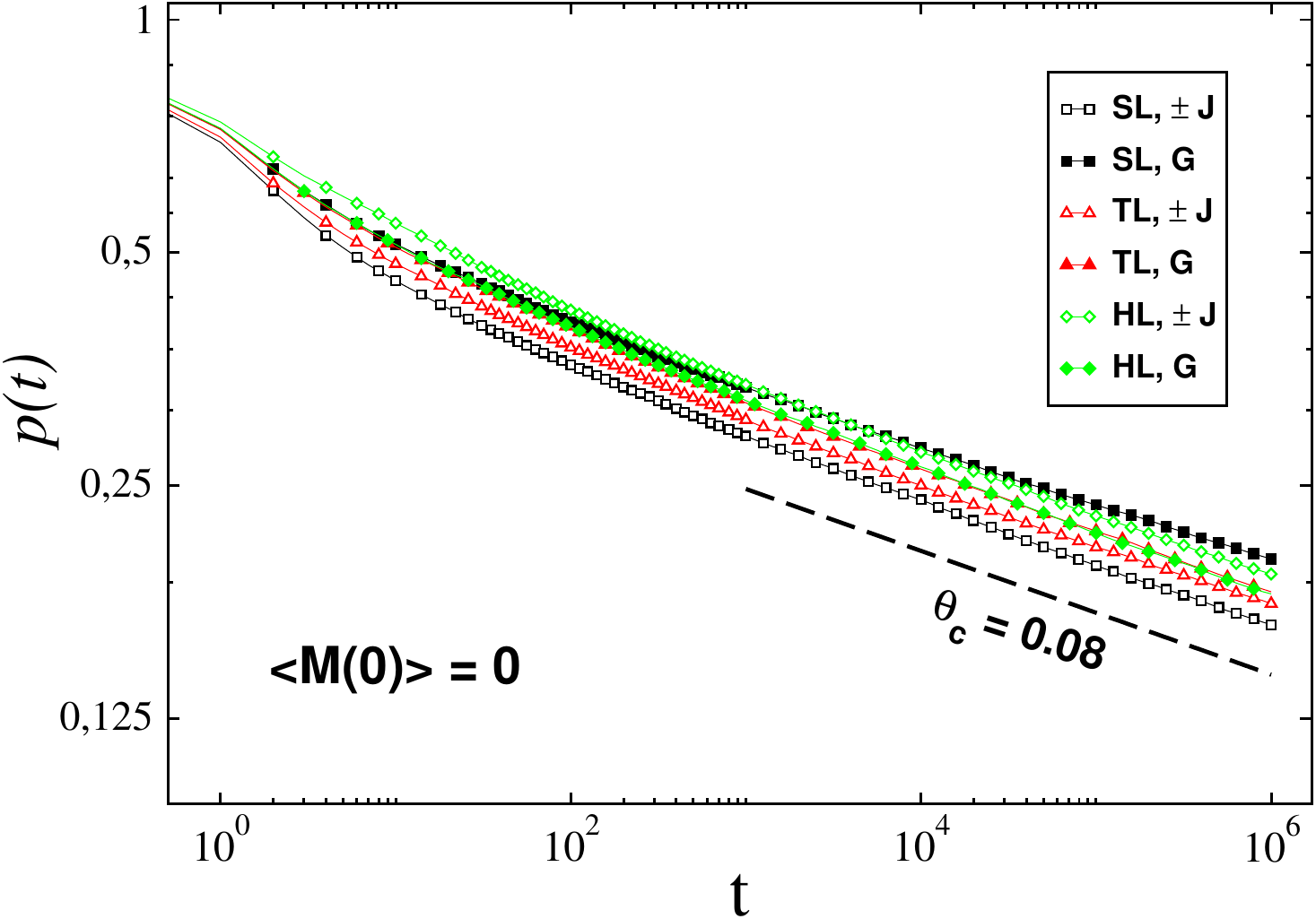}}}
	\caption{Plot of the global persistence probability $p(t)$ vs time $t$, in log-log scale, 
	for a quench to the Nishimori Point $T_{\rm N}$ from infinitely high temperature $T \gg T_{\rm N}$
	initial conditions. Different datasets belong to systems with different lattice geometries and distributions of bond disorder (see the key). The dashed line denotes the power law $p(t) \sim t^{-\theta_c}$, with $\theta_c = 0.08$. The linear size of the system is $L = 128$.}
	\label{fig6}
\end{figure}

\subsection{Global persistence}

\begin{table}[t!]
	\begin{center}
		\begin{tabular}{ c  c  c c }
			\hline
			\hline
			\text{system size} & SL, $\pm J$ & TL, $\pm J$ & HL, $\pm J$ \\
			\hline
			$L = 64$ & $0.48(1)$ & $0.47(2)$ & $0.47(1)$ \\
			$L = 128$ & $0.45(1)$ & $0.44(1)$ & $0.43(2)$ \\
			\hline
			\hline
		\end{tabular}
	\end{center}
	\caption{Values of exponent $a_{\rm c} = \theta_{\rm c} z_{\rm c}$, for a quench from a \textit{sharply prepared} initial state (with magnetization density $M(0) = 0.004$) to the Nishimori Point (NP) at $T = T_{\rm N}$, in the $2d$ frustrated Ising model defined on different geometries [square lattice (SL), triangular lattice (TL), and honeycomb lattice (HL)] with bimodal ($\pm J$) distribution. Different values are obtained by enabling the scaling collapse in Fig.~\ref*{fig7} for the two systems of linear sizes $L=64$ and $128$ defined on each geometry (individual column). The corresponding error bars denote the range of the scaling exponent $a_{\rm c}$ (for fixed $z_{\rm c} = 6$) in which the collapse of the data remains statistically good.}
	\label{tab3}
\end{table}

Let us now discuss the universality features of the global persistence probability $p(t)$~\eqref{pers1} 
after a quench from high $T$ to the NP. In Fig.~\ref{fig6}, we first present data for $p(t)$ vs 
$t$ of an initial condition with $\langle M(0) \rangle = 0$. The idea is that the magnetization for randomly generated configurations of a small system will likely attain a nonzero value, with ensemble averaged magnetization $\langle M(0) \rangle = 0$. The linear size of the system is $L=128$, with an average over around $10^5$ nonequilibrium runs. For large $t$, all the datasets belonging to different lattices and bond disorders in the present figure are consistent with power-law decay, $p(t) \sim t^{-\theta_{\rm c}}$, with $\theta_{\rm c} \simeq 0.080~(05)$. The obtained values from the best power-law fits are listed in Table~\ref{tab2}.

\begin{figure}[t!]
	\centering
	\rotatebox{0}{\resizebox{.85\textwidth}{!}{\includegraphics{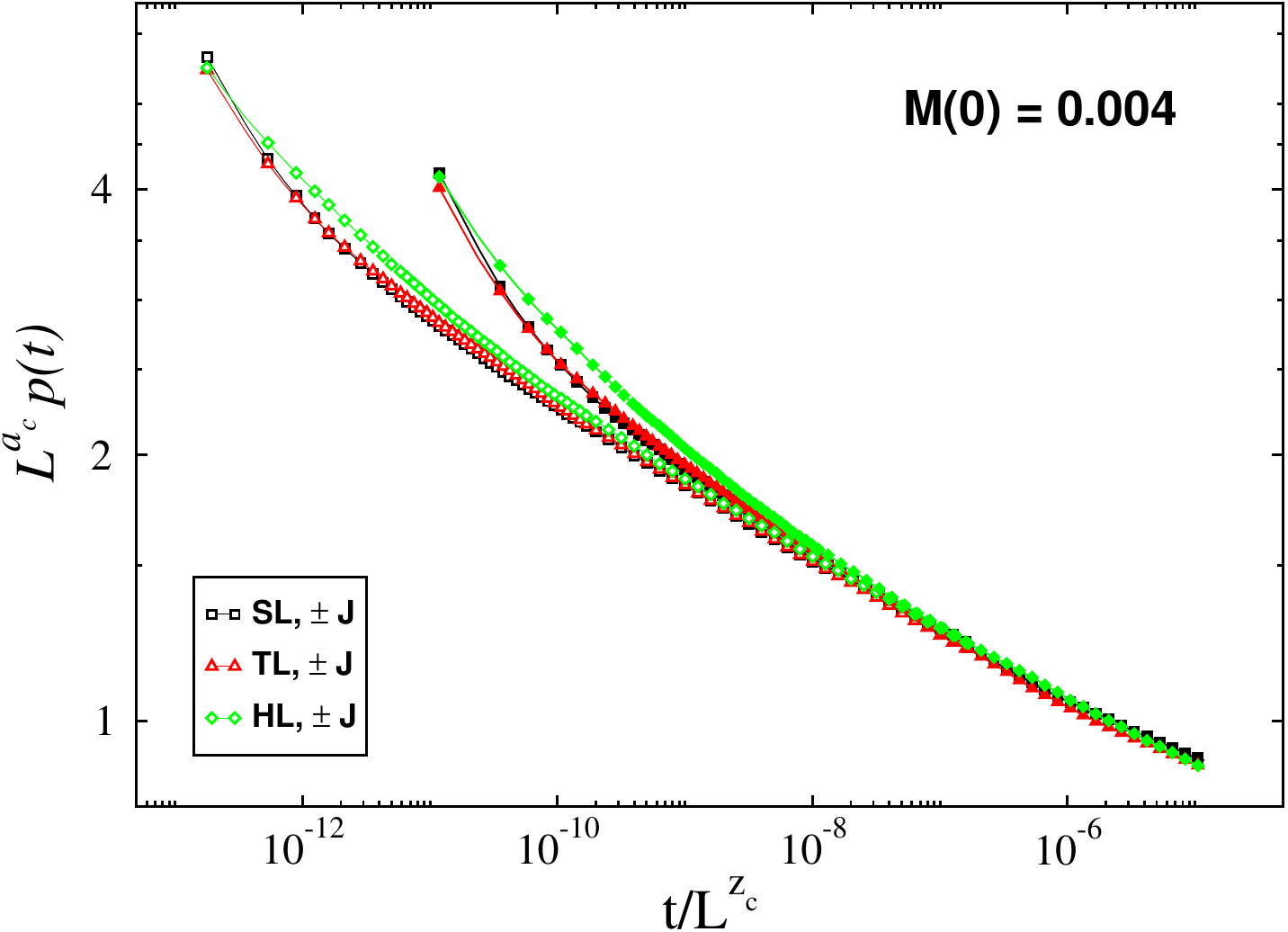}}}
	\caption{Plot of the scaling variables $L^{a_{\rm c}} p(t)$ vs $t/L^{z_{\rm c}}$, in log-log scale, 
	for a quench to the Nishimori point $T_{\rm N}$ from \textit{sharply prepared} initial states 
	with nonzero magnetization density ($M(0) = 0.004$). Different datasets represent systems of 
	different lattice geometries with $\pm J$ quenched-disorder (see the key) and different linear sizes: 
	the linear size of the system for the datasets with empty symbols is $L = 128$, 
	while for the datasets with filled symbols it is $L = 64$.}
	\label{fig7}
\end{figure}

To check what happens to the universality of $\theta_{\rm c}$ when an initial configuration with magnetization density $M(0) \ne 0$ is considered, we also investigate the quench dynamics of a \textit{sharply prepared} initial state. Since $M(0)$ is tiny (we consider $M(0) = 0.004$), finite size effects may eliminate its impact. Therefore, we use a scaling relation~\cite{PhysRevE.67.057102,PhysRevLett.77.3704} obtained by rewriting Eq.~\eqref{pers2} as,
\be
p(t) = L^{-a_{\rm c}} f\left( \frac{t}{L^{z_{\rm c}}} \right)
\label{pers5}
\; ,
\ee
where exponent $a_{\rm c} = \theta_{\rm c} z_{\rm c}$. In Fig.~\ref{fig7}, the scaling variable $L^{a_{\rm c}} p(t)$ is plotted against the scaled time $t/L^{z_{\rm c}}$. Datasets represent systems of linear sizes $L = 64$ and $L=128$, each defined on different lattice geometries with $\pm J$ disorder (the Gaussian case is not shown for brevity). The nonequilibrium averages are performed over $5 \times 10^5$ and $10^5$ runs for $L=64$ and $L=128$, respectively. The values of $a_{\rm c}$ in the scaling variable $L^{a_{\rm c}} p(t)$ are chosen so as to obtain data-collapse, whereas $z_{\rm c}$ is fixed to $z_{\rm c} = 6$ on the time-axis. After preasymptotic time scales, a good scaling collapse is obtained for datasets of both $L$. In fact, the same scaling relation~\eqref{pers5} is obeyed by all systems of different geometries. The obtained values of exponent $a_{\rm c}$ are listed in Table~\ref{tab3}. Upon the increase in $L$, the value of $a_{\rm c}$ seems to decrease. When using $z_c = 6$, a value of $\theta_{\rm c}$ between $0.07 - 0.08$ is obtained from $a_{\rm c} = \theta_{\rm c} z_{\rm c}$, which is also in agreement with $\theta_{\rm c} \simeq 0.080 \pm 0.005$ observed in Fig.~\ref{fig6}. Finally, we conclude that the persistence exponent attains a universal value around $\theta_{\rm c} = 0.080 \pm 0.005$.

In the end, let us also discuss the validity of the relation~\eqref{pers3} based on the Markovian assumption of time-dependent magnetization density $M(t)$. We observed above that the autocorrelation exponent is $\lambda_{\rm c} \simeq 1.32$; while the Fisher exponent is $\eta \simeq 0.18$ at the NP~\cite{PhysRevLett.87.047201}. Using these values in Eq.~\eqref{pers3}, we get $\theta_{\rm c} z_{\rm c} \simeq 0.23$, which is in clear violation of our numerical estimates in Table~\ref{tab3}. Henceforth, the relation~\eqref{pers3} breaks down also for the present frustrated system, suggesting a Non-Markovian behavior of the time-dependent magnetization.

\begin{figure}[t!]
	\centering
	\includegraphics[width=0.98\textwidth]{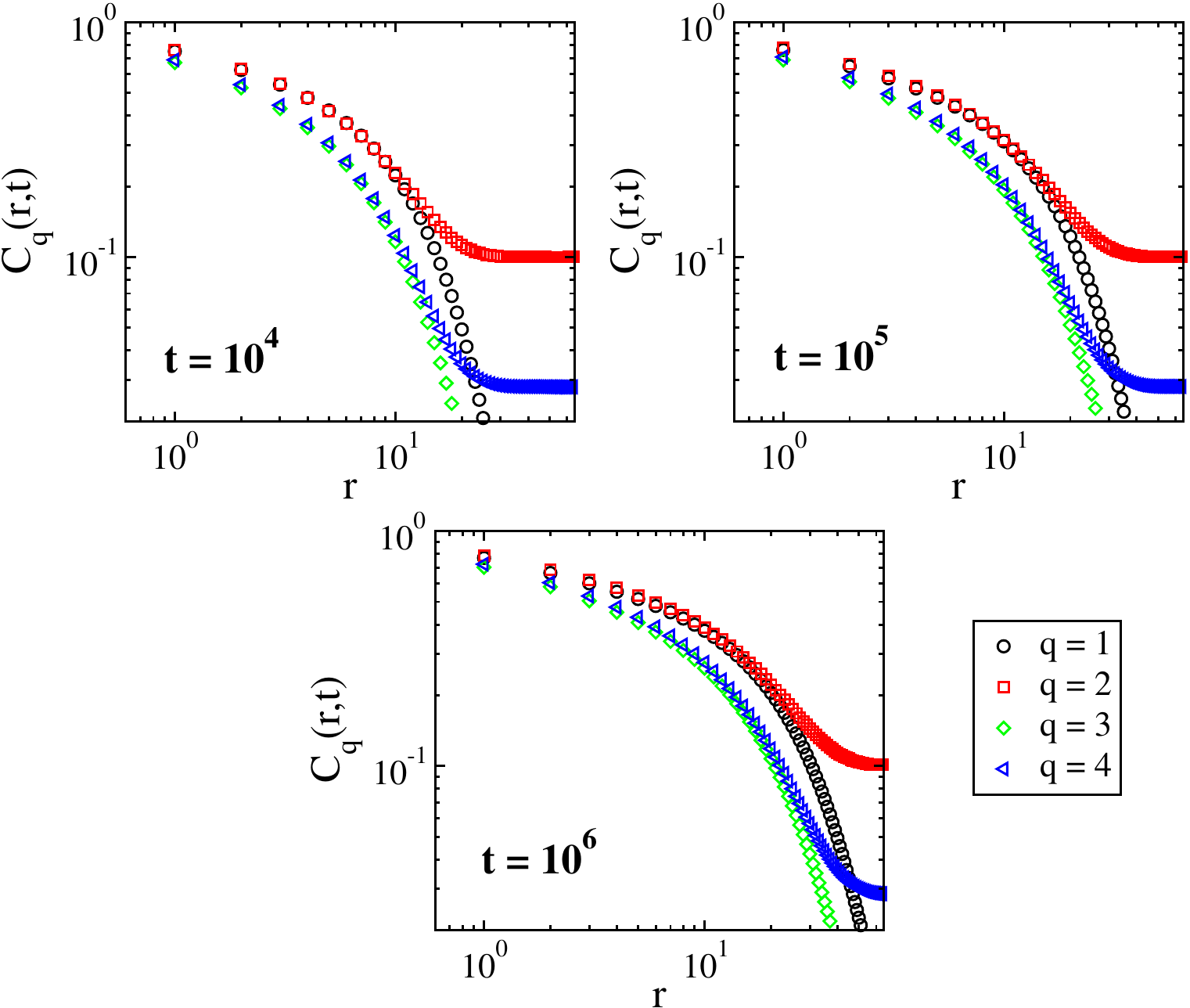}
	\caption{Moments $C_{q}$ vs distance $r$ for different $q$ and at different times 
	written as labels in the panels. The linear size of the system is $L = 128$.}
	\label{fig8}
\end{figure}

\subsection{Multifractality of moments}

We finally discuss the nature of the different moments of the two-point (spatial) 
correlation function during the quench dynamics from the high temperature phase to the NP. 
Let us first define a clean notation for the different moments,
\be
\label{mom0}
C_q(r) = \frac{1}{N} \sum_{i} \overline{ \langle  S_i S_{i+\vec r} \rangle^{q} }; ~~~~~~~~ q = 1, 2, 3, \ldots
\; ,
\ee
in equilibrium, when there is no time-dependence.
The meaning of the averages $\langle (\cdots) \rangle$ and $\overline{ (\cdots) }$ is same 
as the one declared earlier. Along the Nishimori line these moments are equal two by two~\cite{10.1143/PTP.66.1169}. i.e.,
\be
\label{mom1}
C_{2q-1}(r) = C_{2q}(r)
\; ,
\ee
and particularly, at the NP, 
\be
\label{mom2}
C_q(r) \propto \frac{1}{r^{\eta_q}}; ~~~~~~ \eta_1 = \eta_2, \eta_3 = \eta_4, \eta_5 = \eta_6, \ldots
\; .
\ee
The above identities~\eqref{mom1}-\eqref{mom2} are expected to hold, also, during the post quench dynamics to the NP, up to a dynamical length scale unique for each pair of moments. The interesting point to address is whether such length scales corresponding to different pairs of moments have different growth laws.

To answer the above question, we investigate different moments of the correlation function for a system of linear size $L = 128$ with SL geometry. In the calculation of the moments, we have taken nearly $3000$ 
different disorder realizations. For each such realization, $10$ independent runs of the system having different random initial conditions are performed. In Fig.~\ref{fig8}, the identity~\eqref{mom1} is checked for different moments at different times (in MCS) during the nonequilibrium dynamics from the high temperature phase to the NP. For the sake of brevity, we have shown only the first four moments. As it is clear from the different panels of the figure, the moments are equal \textit{two by two} up to some time-dependent length scales. Beyond such scales, the identity~\eqref{mom1} does not hold anymore.

To become more quantitative, we extract the time-dependent length scales from the different moments. Similarly to the average correlation function $C(r,t)$, different moments also have the following scaling behavior during the critical dynamics,
\be
\label{mom3}
C_q(r,t) = \frac{1}{r^{\eta_q}} {\rm e}^{-r/\xi_q(t)}
\; .
\ee
Here the exponent $\eta_q$ is such that $\eta_1 = \eta_2 (\simeq 0.18)$, $\eta_3 = \eta_4 (\simeq 0.26)$, $\eta_5 = \eta_6 (\simeq 0.30)$, and so on. The aforementioned values of $\eta_q$ were calculated in Refs.~\cite{PhysRevLett.87.047201,PhysRevB.65.054425,PhysRevB.73.064410}. $\xi_q(t)$ is the dynamical length scale for different $q$, which can be calculated from the above expression~\eqref{mom3} (see Ref.~\footnote{We have separately checked that different definitions of $\xi_q(t)$ including the one often used in spin glass context (ratio of integrals $\int_{0}^{L/2} dr ~r^2 C_q(r,t)$ and $\int_{0}^{L/2} dr ~r C_q(r,t)$) give equivalent results, apart from a nonuniversal prefactor.} and our previous discussion in Sec.~\ref{S2} on how to estimate the time-dependent correlation length $\xi(t)$).

\begin{figure}[t!]
	\centering
	\includegraphics[width=0.98\textwidth]{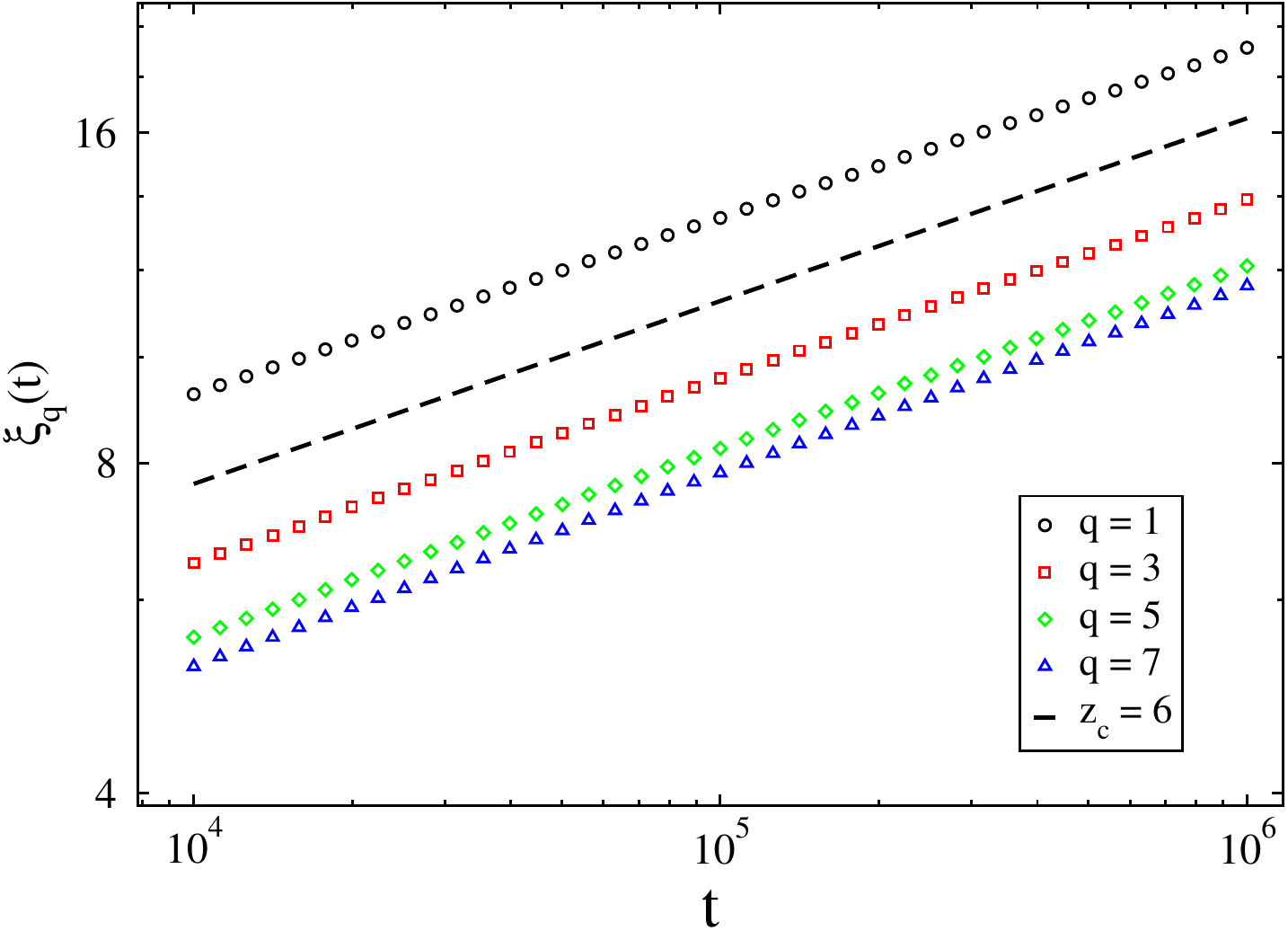}
	\caption{Dynamical length scale $\xi_{q}(t)$ vs time $t$ for different odd values of $q$ (see the key). 
	The dashed line represents the power law, $\xi_q(t) \sim t^{1/z_{\rm c}}$, with $z_{\rm c} = 6$.}
	\label{fig9}
\end{figure}

In Fig.~\ref{fig9}, the length scales $\xi_q(t)$ are plotted against time $t$ for different odd moments $q$. The datasets for different $q$ show algebraic growth on the log-log scale of the figure. The amplitude of $\xi_q(t)$ is smaller for larger $q$, which indicates that at a fixed time $t$ the higher moments fall off in a quicker manner in space; see Eq.~\eqref{mom3}. This is a signature of static multifractality (in equilibrium at NP) during the dynamics. However, the difference in these time-dependent amplitudes remains constant in time, and all the datasets agree with the growth law $\xi_q(t) \sim t^{1/z_{\rm c}}$; with $z_{\rm c} \simeq 6$. This confirms that \textit{there is only a single growth law for all the moments}. At a fixed time during the critical dynamics, different moments of the correlation function maintain the static multifractality, but the associated length scales follow a single growth law in time, i.e, the equilibrium multifractal behavior does not contribute to the dynamics.

\section{Summary and discussion}
\label{S4}

In this paper, we investigated different aspects of the critical dynamics (generated by single spin flip non-conserving order parameter) at the NP in the two-dimensional frustrated Ising models. To the best of our understanding, this is the first study in which the dynamical properties of the multicritical NP are investigated in detail. We calculated the three dynamical critical exponents --- $z_{\rm c}$, $\lambda_{\rm c}$, and $\theta_{\rm c}$. To establish universality of these exponents, we defined the model on different lattice geometries (SL, TL, and HL) with the quenched bond variables drawn from bimodal $(\pm J)$ and Gaussian distributions. We numerically found that after a quench from the high temperature phase to the NP, different dynamical exponents attain their universal values, which confirms that the effects of frustration leave the universality in the dynamical exponents unchanged. This also further consolidates the RG picture of dynamical critical phenomenon in the presence of frustration.

We found that due to systematic errors the value of the exponent $z_{\rm c}$ obtained from power-law fits fluctuates between $6.0 - 6.6$ for different geometries and bond distributions. The effective exponent $z_{\rm eff}$, defined in Eq.~\eqref{effExp}, also endorses the same order of error in $z_{\rm c}$. Notice that the focus of the present work was to establish the universality in the exponents, not the precise determination of the exponent values. In our recent study on the SL with $\pm J$ disorder~\cite{agrawal2023nonequilibrium} (also see Ref.~\cite{Ozeki_1998}), we observed $z_{\rm c} \sim 6$ at late times ($t\sim 10^7$) and large system $(L = 1024)$. Whereas, in the present study, timescales up to $t \sim 10^6$ on $L = 512$ are accessed. Given that, also invoking \textit{Occam's razor}, a universal exponent $z_{\rm c} \sim 6$ is expected in the larger simulations on effectively larger system sizes.

Coming to other dynamical exponents, we obtained a rather precise value of the ratio $\lambda_{\rm c} / z_{\rm c} \simeq 0.22$ from the late time decay of the overlap with the initial configuration, $q_0(t)$. When using $z_{\rm c} \simeq 6$, we get a universal value of $\lambda_{\rm c} \simeq 1.32$. We also investigated the aging during the critical dynamics at the NP, after a quench from the high-$T$ state and the Ising critical point. In both cases, a clean signature of aging in terms of the violation of TTI is observed. Especially, for the critical dynamics of a system quenched from $T = T_{\rm Is}$ , the aging is much slower. The persistence exponent $\theta_{\rm c}$ also produces a universal value around $\theta_{\rm c} \simeq 0.08$.

We also examined the different moments of the spatial correlation function during the dynamics. At the disorder induced fixed points including the NP, the multifractality in the moments is served by the non-integer relation of different magnetic exponents $(\eta_q \ne q \eta_1)$. Recently, the signatures of such multifractality were also reported during the nonequilibrium dynamics of a spin glass system~\cite{januscollaboration2023multifractality}. To probe whether any such behavior arises in our system during the quench dynamics from a high-$T$ state to the NP, we extracted a dynamical length scale from each moment. We found that the length scale for each moment follows the same growth law $\xi_q(t) \sim t^{1/z_{\rm c}}$; $z_{\rm c} \simeq 6$ --- such observation confirms that there is only a single growth law for all the moments. This leads us to conclude that the static multifractality at the NP does not affect the dynamics.

Finally, let us also discuss some possible future directions. In the present work, we confirmed the universality in the dynamical critical exponents at the NP using large-scale numerical simulations. However, for completeness, some analytical understanding of this problem is certainly desirable. Due to the presence of frustration, classical techniques like real space RG method, are not effective to deal with. We believe that semi-analytical approaches like high-temperature series expansions~\cite{PhysRevLett.57.245,Dammann_1993} can be of some use.

\appendix

\section{Nishimori line in phase diagram}
\label{A1}

The model Hamiltonian~\eqref{eq1} in the main text is invariant under the local gauge transformation $S_i \rightarrow S_i k_i,~J_{ij} \rightarrow k_i k_j J_{ij}$; $k_i$ being another Ising variable placed at each site $i$ of the lattice. However, the probability distributions $P(J_{ij})$ in Eqs.~\eqref{eq2}-\eqref{eq3} are not invariant under this gauge transformation. These probability distributions belong to a generic model of distributions~\cite{nishimori2001statistical} defined by
\be
\label{eq5}
P(J_{ij}) = P_0\left(\vert J_{ij}  \vert\right) {\rm e}^{a J_{ij}}
\; .
\ee
The above distribution transforms as: $P(J_{ij}) \rightarrow P_0\left(\vert J_{ij}  \vert\right) {\rm e}^{a J_{ij}k_i k_j}$. The $\pm J$~\eqref{eq2} and Gaussian~\eqref{eq3} distribution functions maintain this form~\eqref{eq5} with $a = - (2J)^{-1} \ln \left( p/(1-p) \right)$ and $a = 1/(J_0 r^2)$, respectively. Here, $p$ and $r = \sigma/J_0$ are the corresponding frustration parameters (see the main text).

Nishimori~\cite{nishimori2001statistical,Nishimori_1980,10.1143/PTP.66.1169} found that when $a = \beta$; $\beta (=1/(k_{\rm B} T))$ being the inverse-temperature, certain thermodynamic properties of the model~\eqref{eq1} can be exactly calculated, e.g., the 
average internal energy, an upper bound on the specific heat, a relation between the moments of the correlation functions. 
Clearly, this condition ($a = \beta$) represents a special curve in the phase diagram relating the temperature ($T$) and frustration parameter ($p$~or~$r$). It is referred to as the \textit{Nishimori line}. For $\pm J$ and Gaussian distributions, 
the Nishimori lines are given by
\be
\label{eq6}
{\rm e}^{-2 \beta J} = \frac{p}{1-p} \, ,  ~~~~~~~ \qquad \beta = \frac{1}{J_0 r^2}
\; , 
\ee
 respectively.

\vspace{0.5cm}

\noindent
{\bf Acknowledgements} The authors acknowledge financial support from the French ANR grant ANR-19-CE30-0014.

\bibliography{ref1}

\end{document}